\documentclass[12pt,preprint,fleqn]{aastex}

\usepackage{graphicx}
\slugcomment{Accepted by {\it The Astrophysical Journal}}

\shorttitle{Neutrino emission from 1ES 1959+650}
\shortauthors{Reimer et al.}

\begin{document}

\title{Neutrino emission in the hadronic Synchrotron Mirror Model: the ``orphan'' 
TeV flare from 1ES 1959+650}

\author{A. Reimer\altaffilmark{1}, M. B\"ottcher\altaffilmark{2} \and S. Postnikov\altaffilmark{2}}

\altaffiltext{1}{Fakult\"at f\"ur Physik \& Astronomie, 
                Lehrstuhl f\"ur Theoretische Physik IV: Weltraum- \& Astrophysik,
                Ruhr-Universit\"at Bochum,
                D-44780 Bochum, Germany; afm@tp4.rub.de}
\altaffiltext{2}{Astrophysical Institute, Department of Physics 
and Astronomy, Ohio University, Athens, OH 45701, USA; mboett@helios.phy.ohiou.edu, postnik@helios.phy.ohiou.edu}

\begin{abstract}
A challenge to standard leptonic SSC models are so-called orphan TeV flares, 
i.e. enhanced very high energy (VHE) gamma-ray emission
without any contemporaneous 
X-ray flaring activity, that have
recently been observed in TeV-blazars (e.g., 
1ES~1959+650).
In order to explain the orphan TeV flare of 1ES~1959+650 
observed in June 2002, the co-called hadronic synchrotron mirror model has 
been developed.
Here, relativistic protons are proposed to exist in the jet, 
and interact with reflected electron-synchrotron radiation of the precursor 
SSC flare.
If the reflector is located in the cloud region, time shifts
of several days are possible between the precursor and the orphan flare.
The external photons, blueshifted in the comoving jet frame,
are able to 
excite the $\Delta (1232)$-resonance when interacting with
protons of 
Lorentz factors $\gamma'_p \sim 10^3$ -- $10^4$.
The decay products of 
this resonance include charged pions
which, upon decay, give rise to 
neutrino production during the orphan flare.
In this paper we 
calculate the expected neutrino emission
for the June 4, 2002, orphan 
TeV flare of 1ES~1959+650.
We compare our results with the recent 
observations of AMANDA-II
of a neutrino event in spatial and temporal coincidence with
the orphan flare of this blazar.
We find that
the expected neutrino signal from the hadronic synchrotron mirror model 
is insufficient to explain the claimed neutrino signal from the direction 
of 1ES~1959+650.
\end{abstract}


\section{Introduction}

Blazars are jet-dominated active galactic nuclei (AGN) that are
viewed close to the jet axis. The observations of high apparent 
luminosities,
short variability time scales and apparent superluminal 
motion of jet components along the jet axis can naturally be explained 
by those components moving with relativistic speeds, and consequently 
Doppler boosting their emitted radiation into the jet direction.
The spectral energy distribution (SED) of blazars generally consists of 
two main
components. The low-energy
one includes continuum emission from 
the radio up to UV or X-ray energies, while the high-energy
hump extends 
into the $\gamma$-rays, in some cases up to TeV-energies. X-ray observations 
reveal a marked flux decline
between both components.

A sub-class of blazars are BL Lac objects, who themselves
are generally 
devided up into high-frequency (HBLs) and low-frequency peaked
BL Lacs 
(LBLs), depending on the peak frequency of their low-energy component,
with a smooth transition between both classes.
Up to now, only HBLs have 
been detected in the TeV energy range, in both,
flaring and quiescent 
states. The TeV source catalog of AGN contains
so far only six objects:
Mrk~421 \citep{punch92}, Mrk~501 \citep{quinn96}, PKS 2155-314 
\citep{chadwick99,HESS05}, 1ES~2344+514 \citep{catanese98},
1H~1426+428 \citep{horan02}, 1ES~1959+650 \citep{kajino99,holder03}),
have now been detected at VHE $\gamma$-rays ($> 150$~GeV) by 
ground-based air \v Cerenkov telescopes. They exhibit variability 
at all wavelengths and on various time scales. 

Depending on the nature of the relativistic particle content in AGN jets, 
electron-positron pairs or electron-proton matter, the high-frequency 
(X-ray -- $\gamma$-ray) emission could either be produced via Compton 
upscattering of low frequency radiation by the
same electrons responsible 
for the synchrotron emission \citep[for a recent review see, e.g.,][]{boettcher02}, 
or due to hadronic processes initiated by relativistic protons co-accelerated 
with the electrons \citep[for a recent discussion see, e.g.][]{muecke03}. 
The difficulty
in understanding the primary jet launching mechanism and 
constraining the jet composition from general energetics considerations
currently leave both leptonic and hadronic models open as viable possibilities.

In the framework of leptonic jet models, the TeV emission from HBLs is 
often successfully
modelled by inverse Compton scattering of internal 
photon fields (i.e. the e$^-$ synchrotron radiation), while external 
target radiation fields,
owing to the weak accretion disks of BL Lac 
objects, are in general considered
negligible. These so-called SSC models 
predict that any flaring activity
at TeV energies should leave its imprints 
in the low energy synchrotron component
as enhanced emission. 
This is in striking contrast to the recent observation of
\cite{kraw04} 
of an orphan TeV flare seen in the Whipple light
curve of the TeV blazar 
1ES~1959+650 during a multiwavelength campaign
in May and June 2002. 
In May 2002 the blazar displayed a TeV-flare that was accompagnied by 
X-ray activity (observed by RXTE). This outburst (called SSC-flare in 
the following)
can be modelled using an SSC code \citep{kraw04}. During 
the subsequent (well
sampled) decline of the X-ray flux over the following 
month, a second TeV flare, on June 4, i.e. $\sim 15$~days after the initial 
one, was observed. No significant spectral changes has been observed with 
Whipple
between the SSC- and the orphan flare \citep{daniel05}.
Only very 
moderate $\lesssim 0.1^{\rm mag}$
flaring activity in the R and V bands 
was observed during this time. The lack of enhanced X-ray activity during 
the second TeV-flare
can not be explained by purely leptonic SSC blazar 
jet models.
The occurence of orphan TeV-flares during
multifrequency 
campaigns has also recently been claimed from Mrk 421 \citep{blaze05}.
Here, the orphan flare lagged an SSC-flare by 1 -- 2~days.

Motivated by the observations of the 1ES~1959+650 orphan flare, \cite{bo05} 
has developed
a hadronic model which considers its occurence in connection 
with a correlated
SSC X-ray and TeV-flare. In this model, the secondary 
orphan TeV-flare is explained by $\pi^0$-decay ($\pi^0 \to 2\gamma$) 
$\gamma$-rays as a result of $p\gamma \to \Delta \to p\pi^0$ pion production 
in the $\Delta~(1232)$-resonance from relativistic
protons of energy 
$10^3$ -- $10^4$~GeV interacting with the primary synchrotron-flare photons
that have been reflected off clouds 
located at a few pc above the accretion disk.
The reflected 
external photons appear Doppler blue shifted in the jet-frame, which 
substantially lowers the required proton
energy to overcome the threshold 
for $p\gamma$ pion production as compared to the
standard hadronic-jet 
scenario based on synchrotron target photons \citep[e.g.,][]{atoyan}.
This relaxes the requirements on the proton acceleration mechanism.

The branching ratio for the charged pion production channel
$p\gamma \rightarrow 
\Delta^+ \rightarrow n\pi^+$ is $\sim 33.3$~\% \citep[e.g.,][]{muecke00}, and 
the resulting $\pi^+$ decay according to
$\pi^+ \to \mu^+ + \nu_{\mu}$ is followed 
by $\mu^+ \to \overline{\nu_{\mu}} + e^+ + \nu_e$, thereby producing 3 neutrinos
per $\pi^+$. Recently, the AMANDA collaboration has reported two neutrino 
events within a 66~day window
around the 1ES~1959+650 orphan flare of June 2002, 
where one occured exactly during the
orphan outburst and the second 31 days 
later \citep{nu_event,nu_event2}. Their statistical significance,
however, 
can not be reliably estimated since these events 
do not obey the original criteria for the blind analysis.

If the produced photons (and pairs) do not initiate electromagnetic cascades, 
the
observed $\pi^0$-decay $\gamma$-ray energy is in general related to the 
neutrino energy by
$E_\gamma \approx 2 E_\nu$ (e.g. \cite{muecke99}). The 
radiative output $L_{\pi^0}$ from the $\pi^0$-decay channel
is connected 
to the total neutrino output, $L_{\nu}$, by $L_{\pi^0} \approx (2/3) L_{\nu}$, 
i.e., the $\pi^0$-decay photon and muon neutrino fluxes are roughly equal in 
the absence of neutrino oscillations. Thus, for an orphan flare flux of 
$\nu F_\nu (600 \, {\rm GeV}) \sim 3 \times 10^{-10}$~erg~s$^{-1}$~cm$^{-2}$ 
as measured by Whipple
\citep{kraw04}, the expected $\nu_\mu$-flux at Earth 
is roughly
$E_{\nu} F_{\nu_\mu} (E_\nu \sim 300 \, {\rm GeV}) \sim \frac{1}{3} 
3 \times 10^{-10}$~erg~s$^{-1}$~cm$^{-2}$ for an expected
flavor ratio after 
$\nu$-oscillations of $\nu_\mu : \nu_e : \nu_\tau = 1 : 1 : 1$ if photon 
absorption
in the cosmic background radiation field is not considered. This 
$\nu_\mu$-flux value of
$F_{\nu_\mu} \sim $~a~few~$\times 
10^{-10}$~s$^{-1}$~cm$^{-2}$ is rather a lower limit, and is comparable to 
the atmospheric
$\nu$-background for a resolution element of $~0.1\degr \times 
0.1\degr$.

This paper presents the calculation of the expected neutrino emission during the
1ES~1959+650 orphan flare of June 2002 using the steady-state Monte Carlo jet 
emission
code of \cite{muecke01,muecke03}, modified to allow for the hadronic 
synchrotron mirror (= HSM) model scenario. This code uses the SOPHIA Monte Carlo 
program
\citep{muecke00} for secondary particle production and decay.
A 
corresponding fully time- and angle-dependent calculation will be presented 
elsewhere \citep{Postnikov05}.
In the last section, we will present a 
discussion on the detectability of this
$\nu$-flux with current and near-future 
observatories.

\section{\label{SED}Photon emission}

In order to calculate the $\gamma$-ray emission in the framework of the HSM model
we modified the synchrotron proton blazar (SPB) model code (see \cite{muecke01} 
for a description of the
SPB model implementation) to account for the reflected 
jet synchrotron emission as an external target photon field.
Its energy density 
in the jet-frame is given by (\cite{bo05})
\begin{equation}
{u'}_{\rm Rsy} \sim {{\tau_m \, \Gamma^2 \,  d_L^2 \over 4 \pi R_m^2 c} \, 
\nu F_{\nu} ({\rm sy})}
\label{u_Rsy}
\end{equation}
if the reflecting cloud is located in a direction close to our line of sight.
Here and in the following, primed quantities are considered in the jet comoving 
frame.
$\tau_m$ is the reprocessing optical depth of the mirror, $R_m \approx 
2 \Gamma^2 c \Delta t_{\rm obs}$ the distance of the reflector from the 
accretion disk, estimated through the observed time delay
$\Delta t_{obs}$ 
between the primary and the orphan flare, $\Gamma \approx D$ the bulk 
Lorentz factor with
$D$ the bulk Doppler factor, $d_L$ the luminosity 
distance of the blazar, and $\nu F_{\nu} ({\rm sy})$
the observed 
synchrotron luminosity during the primary $\gamma$-ray flare.
This 
photon field, simplified as an isotropic distribution in the jet frame, 
serves as target for hadronic proton-photon pion production, electromagnetic
proton-photon collisions and in principle electromagnetic cascading. 
Considering a beam of photons in the jet frame, instead, increases the 
resulting photon and neutrino luminosities by less than 20~\%, indicating
that our simplifying assumption of source photon isotropy has only a minor
effect on the final results. We postpone
a corresponding calculation to a 
later work \citep{Postnikov05}.

We base the determination of ${u'}_{\rm Rsy}$ for the 2002 1ES~1959+650 
orphan flare on the SSC modelling of the primary flare SED
on May 7 2002 
as shown in \cite{kraw04}, which gives $B'=0.04$~G for the magnetic field 
strength, $D=20$, comoving blob size $R' \sim 10^{16}$~cm and 
$\Delta t_{\rm obs} = 15$~days. For a redshift $z = 0.047$ the blazar 
lies at a distance of $d_L \approx 210$~Mpc.
\cite{bo05} demonstrated 
that a Compton optical depth of $\tau_m = 0.1$ for the cloud is in agreement
with a BL~Lac classification of 1ES~1959+650 concerning its emission line 
luminosity, and typical cloud sizes and densities
in the pc-scale vicinity 
of the central engine. We inject a proton spectrum $N'_p \propto 
\gamma_p^{'-\alpha_p} \exp(-\gamma'_p/\gamma'_{p,\rm max})$ with 
$D = \gamma'_{p,\rm min} \leq \gamma'_p \leq \gamma'_{p,\rm max}$
and $\alpha_p = 2$. Our calculations take into account losses
by 
$p\gamma$ pion production and Bethe-Heitler pair production (while 
proton-, muon and pion synchrotron
losses are negligible). For 
$\pi$-production we use the SOPHIA Monte Carlo event generator 
\citep{muecke00} which
takes into account the full cross section 
modelled by resonance excitation and decay, direct single pion
production, diffractive and non-diffractive multiparticle production.

For the parameters given above
we find the $\gamma\gamma \to e^+e^-$ 
optical depth in the blob is too small to initiate any cascading of the
produced $\pi^0$-decay radiation. The escaping $\pi^0$-decay $\gamma$-rays 
suffer, however, from absorption
in the cosmic background radiation field 
during their propagation to the Earth. The recent background model
of 
\cite{P04} (= Primack04) promotes only minor absorption, while the 
phenomenological shape used in \cite{aha03} (= Phigh; essentially
scaled up and smoothed using the \cite{P01} calculation) gives an upper 
limit thereof.
A model intermediate in absolute background flux, but 
with a different spectral shape, is
provided by \cite{MSbaseline} 
(=MSbaseline).

\begin{figure}[t]
\includegraphics[height=13cm]{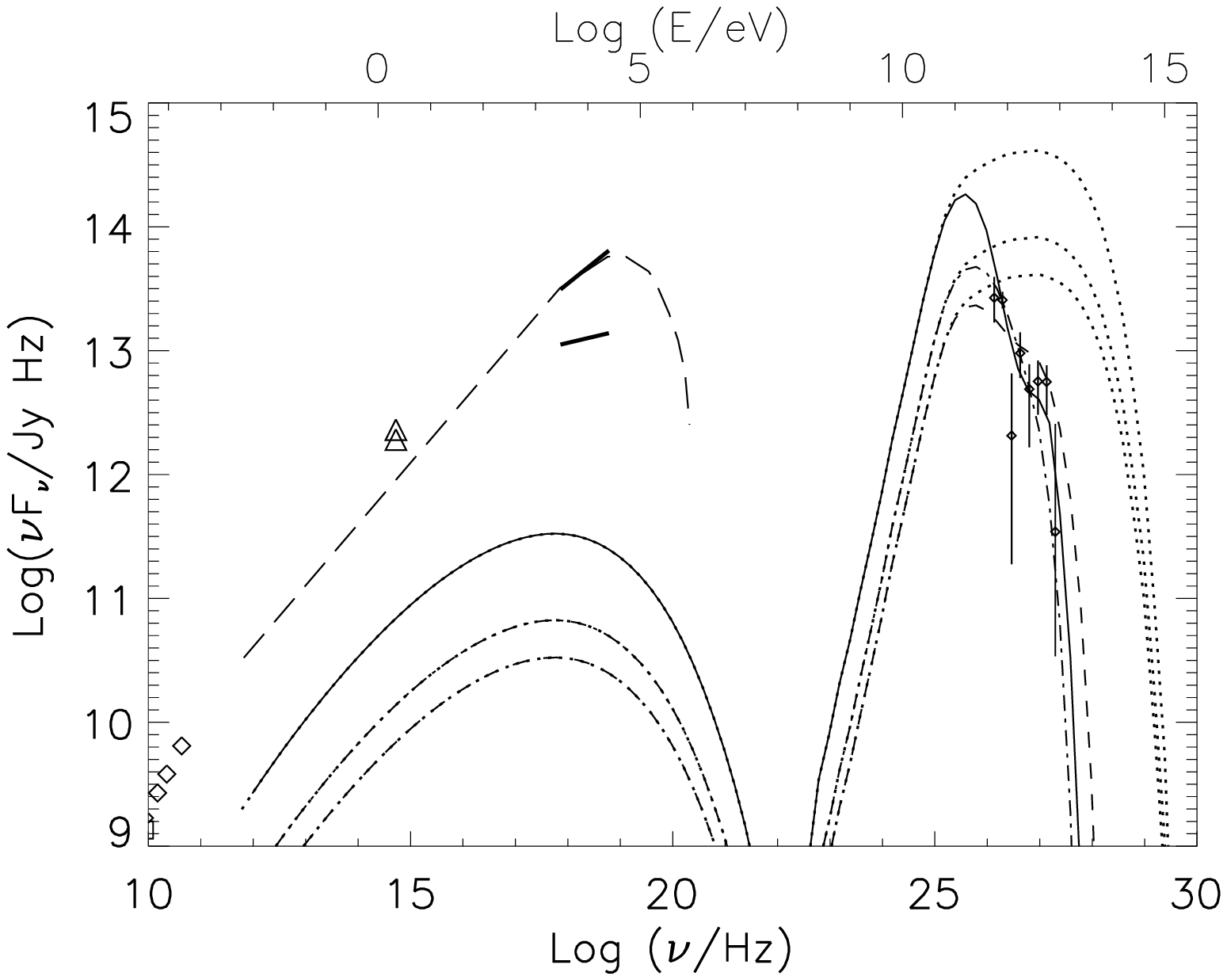}
\caption{Broadband spectral energy distribution of 1ES~1959+650. 
The long-dashed line represents the electron synchrotron component
of the primary TeV flare as modelled by \cite{kraw04} and scaled to account for the May X-ray data. 
The RXTE data of the low and high state are represented by the heavy solid lines.
The dotted line shows the calculated de-absorbed $\pi^0$-decay $\gamma$-ray spectrum
for an $\alpha_p=2$ proton injection spectrum,
while the solid, dashed and dashed-dotted lines refer to the case where absorption in the cosmic background 
radiation field has been taken into account, using the background model of \cite{aha03} (=Phigh), 
\cite{P04} (=Primack04) and \cite{MSbaseline} (=MSbaseline), respectively.
The corresponding predicted secondary $e^+$ synchrotron emission following
$\pi^+$ decay in the IR-to-X-ray range is indicated by the same line styles for each background model.
The parameters are:
$B'=0.04$~G, $R'=10^{16}$cm, $D=20$, $\tau_m=0.1$, $\Delta t_{\rm obs}=15$days, $\gamma'_{p,\rm min}=D$,
$\gamma'_{p,\rm max}=10^4$, $u'_p = 0.45$, 0.045 and 0.09 erg/cm$^3$ for the Phigh, Primack04 and MSbaseline model, respectively. Data are from \cite{kraw04} and \cite{daniel05}.}
\label{SEDfig}
\end{figure}

\begin{figure}[t]
\includegraphics[height=13cm]{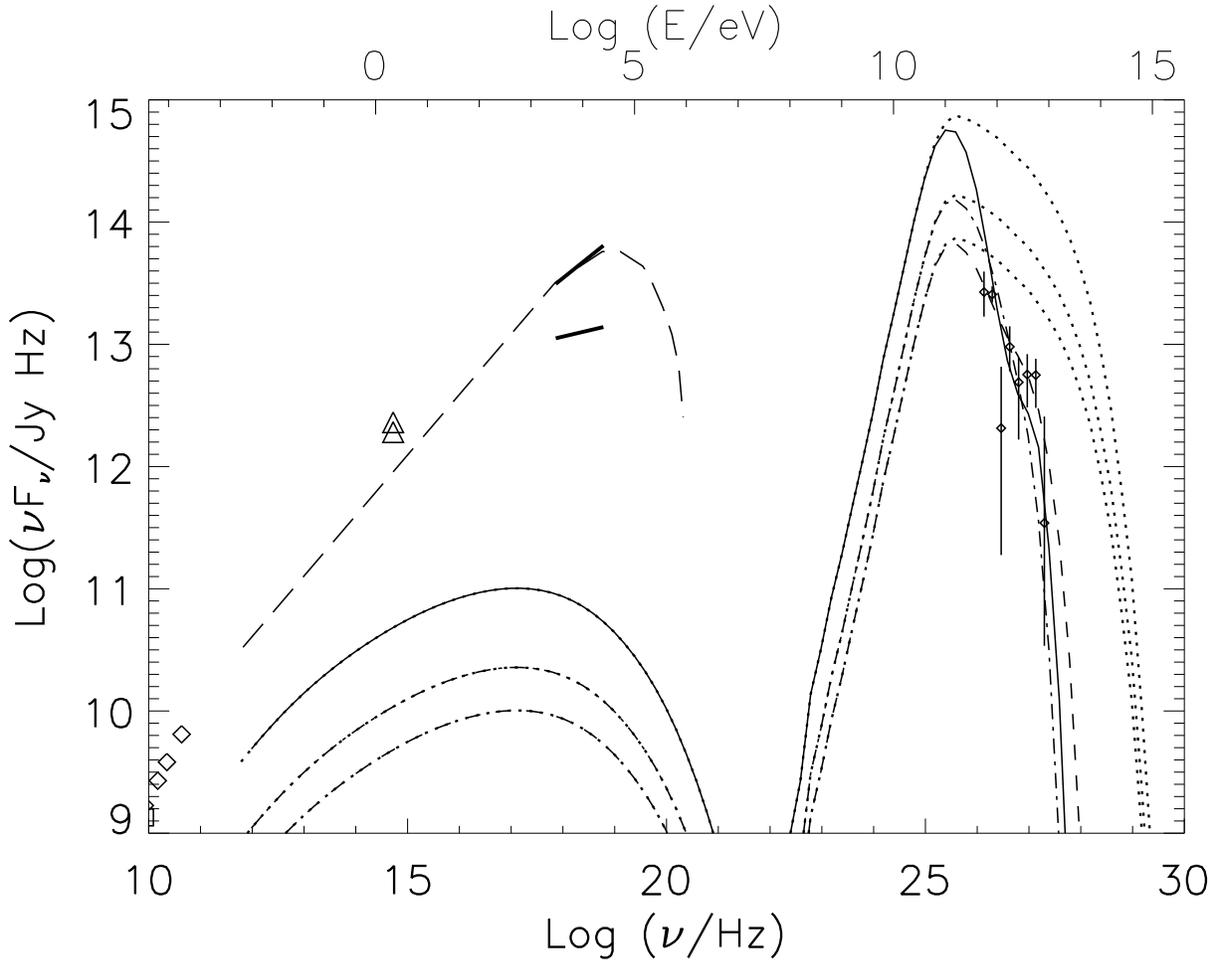}
\caption{Same as Fig.~\ref{SEDfig} but with $\alpha_p=2.5$ and $u'_p = 1.2$, 0.12 and 0.27 erg/cm$^3$ 
for the Phigh, Primack04 and MSbaseline model, respectively.}
\label{SEDfig2}
\end{figure}

Fig.~\ref{SEDfig} shows the resulting spectral energy distribution that 
provides a reasonable representation of the
observed Whipple orphan flare 
spectrum of 1ES~1959+650 as $\gamma$-rays from the $\pi^0$-decay for a maximum
proton Lorentz factor of $\gamma'_{p,\rm max} = 10^4$ with a $\alpha_p=2$ proton 
injection spectral index and using various cosmic background models.
A total 
energy density of relativistic protons of $u'_p = 0.045$~ergs~cm$^{-3}$ is 
required for the Primack04-model, while a fit
using the maximal absorption 
background model (Phigh) needs a factor 10 more injected protons. A variation 
of the injected proton spectral index $\alpha_p$ to steeper values provides 
slightly
better fits when using the Primack04-model, while the MSbaseline 
model would prefer harder injection spectra
(see Fig.~\ref{SEDfig2}).
A  change of $\alpha_p$ from 2 to 2.5 requires a general increase of $u'_p$ 
by a factor 1.7 in order to fit the data.

In the giant resonance region, $\sim 33$~\% of all produced pions are charged 
\citep{muecke00}.
The charged pions decay, thereby producing copious positrons, 
which will primarily loose their energy
through synchrotron radiation before 
potentially leaking out of the emission region. Because of the low field 
strength $B'_G$ (in Gauss) only a fraction 
$f = w_{\rm obs}^{\pi^0}/w_{\rm obs}^{\rm esyn}$
of the positron energy will be lost during the $\pi^0$-decay $\gamma$-ray flare, 
where $w_{\rm obs}^{\pi^0}\approx R_m/(8\Gamma^4 c)$ is the (observer-frame) 
duration of the $\gamma$-ray flare \citep{bo05} and $w_{\rm obs}^{\rm esyn}
\approx 1.4\cdot 10^7 (\Gamma B_G^{'2} \gamma'_p)^{-1}$~sec
is the 
observer-frame synchrotron loss time scale of the secondary positrons. 
The mean positron energy
is $\gamma'_e \approx 55 \gamma'_p$ \citep{muecke99}. 
Thus, we weight
the synchrotron radiation of the pairs produced by each 
$\gamma'_p$ with $f$. In Figs.~\ref{SEDfig}, \ref{SEDfig2}
we also present the resulting
secondaries' synchrotron radiation which lies 
$\sim 2-3$ orders of magnitude below the optical R-band data points 
\citep[from][]{kraw04}.

\section{\label{nuprod}Neutrino production}

\begin{figure}[t]
\includegraphics[height=13cm]{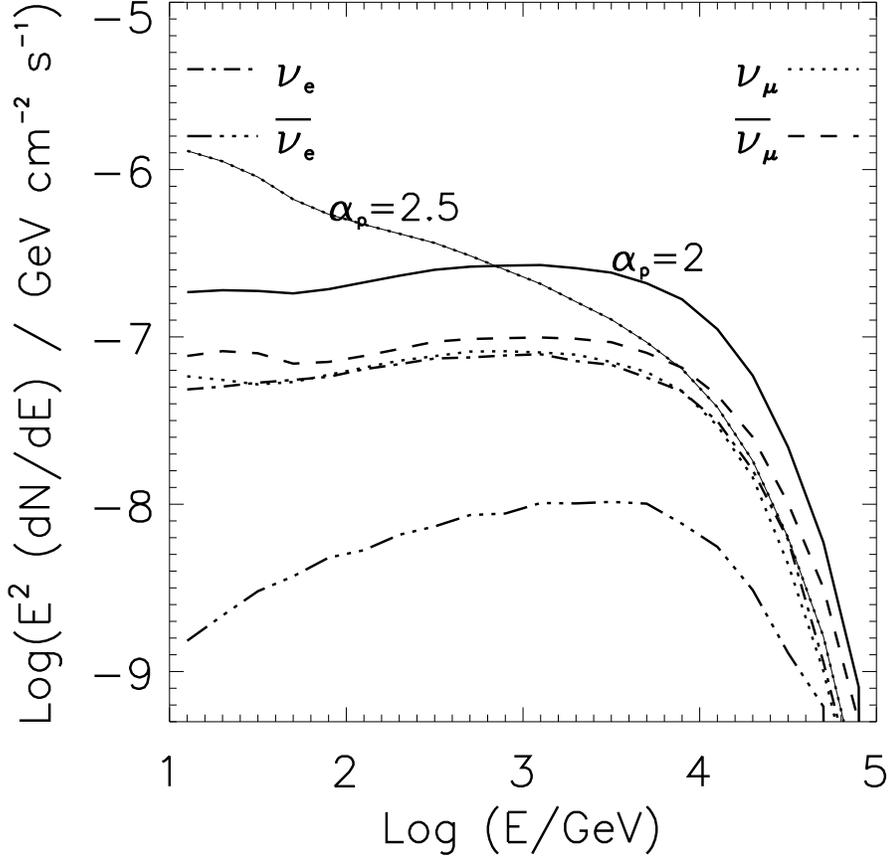}
\caption{Predicted neutrino spectrum of 1ES~1959+650 during the "orphan" flare.
The dashed, dotted, dashed-dotted and dashed-triple dotted lines refer to a $\alpha_p=2$ injection
spectrum.
The solid lines represent the total neutrino ($\nu_\mu+\bar\nu_\mu+\nu_e+\bar\nu_e$) flux
for a $\alpha_p=2$ and $\alpha_p=2.5$ injection spectrum and using the low photon absorption
model \citep{P04}. The use of a high photon absorption model (Phigh) leads to a factor 10 higher $\nu$-flux.
Neutrino oscillations nor $\beta$-decay neutrinos has not been taken into account here.
}
\label{nu_spec}
\end{figure}

In hadronic models neutrino production occurs simultaneously to 
$\pi^0$-decay $\gamma$-ray production.
The main neutrino production 
channel is
through the decay of charged pions, e.g., $\pi^\pm\to\mu^\pm
+\nu_{\mu}/\bar{\nu_{\mu}}$ followed by $\mu^\pm\to e^\pm +
\nu_e/\bar{\nu_e} + \nu_{\mu}/\bar{\nu}_{\mu}$. The neutrinos
escape without further interaction. Fig.~\ref{nu_spec}
shows the 
predicted neutrino emission in steady-state from 1ES~1959+650
during the orphan flare in 2002.
The photon-hadron interactions 
take place predominantly in the resonance region, where
$\pi^-$ and thus $\bar{\nu_e}$ production is suppressed.  

The neutrino production spectrum depends in general on the injection 
proton
spectrum modified by interactions and energy
losses, and the 
spectrum and density of target photons. For the 1ES~1959+650 flare
the photopion production rate approximately
follows a broken power 
law, $(\dot{\gamma_p}' / {\gamma_p}')_{\pi} \equiv t_{\pi}^{'-1} 
\propto \gamma_p^{'1.2}$
for proton energies below $\sim 10^{3}$~GeV, 
and $t_{\pi}^{'-1}
\propto \gamma_p^{'0.5}$ above due to the
break in the target photon spectrum (see Fig.~\ref{SEDfig}).
The cutoff in the $\nu$-spectrum is a consequence of the 
injection proton spectrum turnover at $10^4$~GeV.

In addition to the neutrinos from the meson decay chain, there
will 
be a small contribution of $\bar{\nu}_e$ from the decay of neutrons
(not shown in the figure) that have been produced. For $R'=10^{16}$~cm, 
only neutrons of energy $\sim 300$~GeV will decay inside the production 
region.
High-energy neutrons leaving the source into the observer's
sight line will decay and produce highly beamed neutrinos. Those
neutrinos also appear to originate from the AGN.
In addition, there 
will be another contribution from neutrons
decaying close enough to 
the system to appear as a point source to
the observer at cosmological
distances. In 
the following we disregard this $\nu$-contribution since we aim mainly 
for the expected muon neutrino flux arriving at Earth.
We do not consider any additional contribution from
escaping cosmic rays interacting while 
propagating through the
cosmic microwave background radiation.

For an $\alpha_p = 2$ proton spectrum and using the Primack04 background 
model we find the expected integrated neutrino flux at $E_{\nu} > 100$~GeV 
to be $2.2\cdot 10^{-9}$cm$^{-2}$ s$^{-1}$ with a neutrino flavor ratio of
$\nu_\mu:\nu_e:\nu_\tau=2:1:0$. Neutrino oscillations during intergalactic 
propagation
leads to a $\nu_\mu:\nu_e:\nu_\tau=1:1:1$ flavor ratio upon 
arrival at the Earth. The muon neutrino
flux arriving at Earth can thus 
be estimated to $7.9 \times 10^{-10}$cm$^{-2}$~s$^{-1}$ ($>100$~GeV).
The absolute neutrino flux depends on the $\gamma$-ray absorption in the 
cosmic background radiation field and on the unknown proton injection 
spectral index. We estimate the uncertainty from different cosmic 
background models to not more than a factor of 10 by using two extreme 
background models (see Sec.~\ref{SED}), namely \cite{P04} as a low and 
\cite{aha03} as a high absorption model, to fit the $\gamma$-ray data.
When varying the injection proton spectrum from $\alpha_p = 2$ -- 2.5, 
the
resulting $\nu$-flux changes by not more than a factor 2.

\section{\label{nuevents}Neutrino event rates and detectibility}

Within the HSM model a neutrino flare is expected simultaneously
to the TeV orphan flare and with an duration $w_{\rm obs}^{\pi^0}
\approx R_m/(8\Gamma^4 c)$
\citep{bo05} that is comparable to the 
$\gamma$-ray outburst.
For the 2002 orphan flare of 1ES~1959+650, 
$w_{\rm obs}^{\pi^0} \sim 1/3$~hr.
We calculate the expected 
$\nu_\mu$ event rate from this flare by 
\begin{equation}
\dot n_{\rm AGN} = \int_{E_{\nu,min}}^{E_{\nu,max}} dE_\nu 
\frac{d\Phi_{\nu,{\rm AGN}}}{dE_\nu} A_{\rm eff}^{\nu}(E_\nu,\delta)
\end{equation}
where $d\Phi_{\nu,{\rm AGN}}/dE_\nu$ is the differential $\nu_\mu$-flux 
and $A_{\rm eff}^{\nu} (E_\nu,\delta)$ the effective area for 
$\nu$-detection which
is dependent on the neutrino energy and the
declination $\delta$ of the AGN. For the determination of the
expected AMANDA-II event rate the compilation of the effective
$\nu$-detection area as published in \cite{AMANDA} and \cite{Ahrens} 
is used, while
for IceCube we assumed a factor 100 improvement of 
the effective area with
respect to AMANDA-II.

For an $\alpha_p = 2$ proton spectrum, we anticipate a $\nu_\mu$-rate 
in AMANDA-II of
$\dot n_{\rm 1ES \, 1959+650} \sim 9.6 \times 
10^{-4 \ldots -5}$~events/hr emitted during the 1ES~1959+650 orphan 
flare in June 2002.
This is several orders of magnitude below the 
observed events by AMANDA-II.
Even for IceCube with an anticipated 
factor 100 improvement of the effective area, the detection of such 
flares is only possible if the signals
of many flaring events are co-added.

\begin{figure}[t]
\includegraphics[height=13cm]{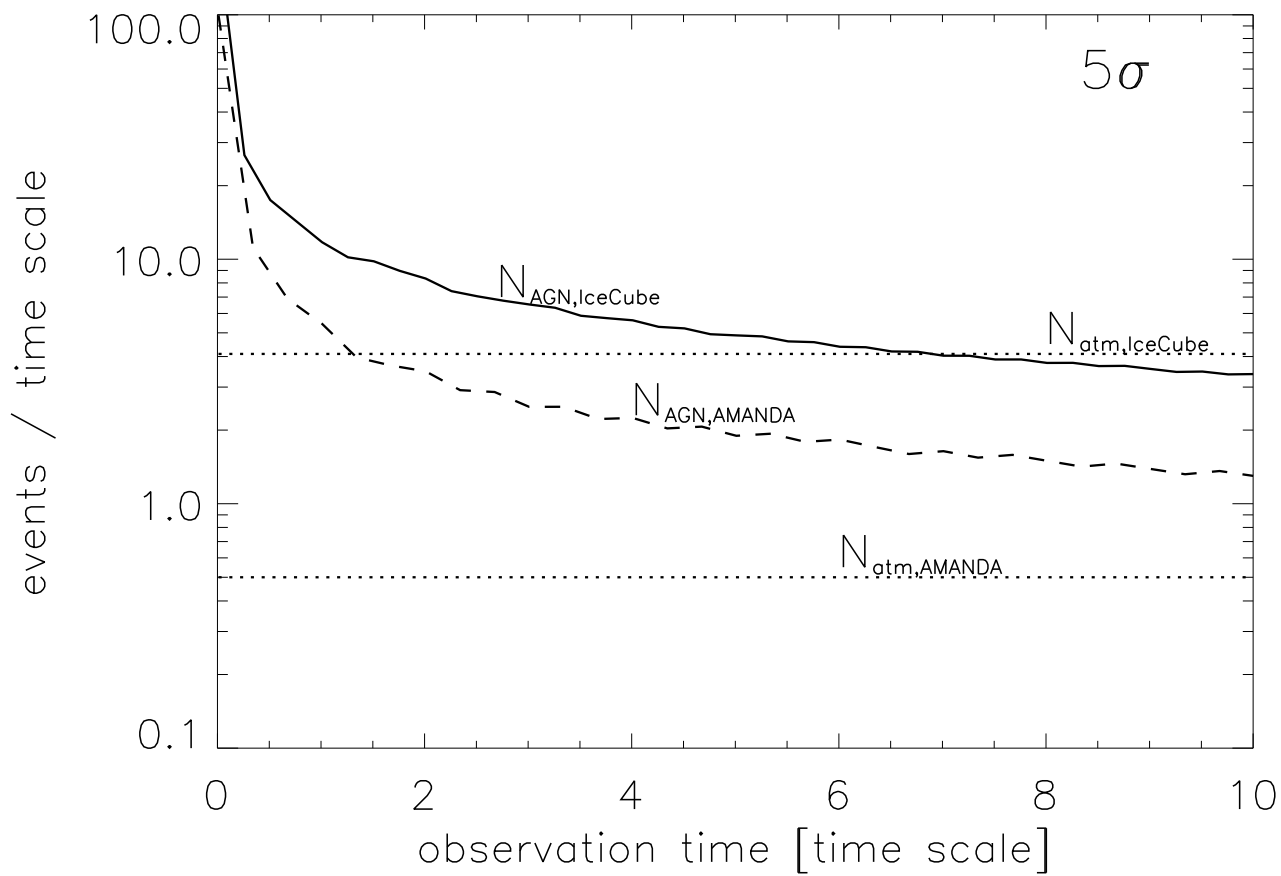}
\caption{The required AGN neutrino event rate in AMANDA-II (dashed line) and IceCube (solid line)
for a $3\sigma$ excess per ($2.5\degr\times 2.5\degr$) resolution element
as a function of the observing time, measured {\it in months for IceCube} and 
{\it in years for AMANDA-II}.
The dotted and dashed-dotted line represents
the event rate from the atmospheric background for IceCube and AMANDA-II, respectively.
For an AGN neutrino rate of $\sim$(0.8-8)/year (AMANDA-II) and $\sim$(7-70)/month (IceCube)
an observing time of $\sim0.5\ldots >10$ years and $<$2 months, respectively, would be needed to produce a 
5$\sigma$ signal for a source flaring throughout the observation time.
}
\label{sig}
\end{figure}

For a meaningful signal of AGN neutrinos we need to incorporate the 
nearly isotropic background, predominantly from atmospheric neutrinos, 
which significantly aggravates the challenge for detections of events
of extraterrestrial origin at GeV -- PeV energies.
The expected event 
rate $\dot n_{\rm atmos}$ from the atmospheric $\nu$-background
is 
obtained by
\begin{equation}
\dot n_{\rm atmos} = \int_\Omega d\Omega \int_{E_{\nu,min}}^{E_{\nu,max}} dE_\nu 
\frac{d\Phi_{\nu,\rm atmos}}{d\Omega dE_\nu} A_{\rm eff}^\nu(E_\nu,\delta)
\end{equation}
with $d\Phi_{\nu,{\rm atmos}}/d\Omega dE_\nu$ the differential atmospheric 
$\nu$-flux
in a solid angle element $d\Omega$ and energy element $dE_\nu$. 
We use a recent compilation of the atmospheric $\nu$-flux
of \cite{spiering05} for the present work.
AMANDA-II's capabilities in reconstructing $\nu$-events 
is estimated to $\sim 2.5\degr$ \citep{spiering05}, 
which corresponds to a 
resolution element of $\sim 2 \times 10^{-3}$~sr.
The number of atmospheric 
$\nu$-events ($>100$~GeV) per year and resolution element for AMANDA-II is 
then
0.2 and 0.8 for vertical and horizontal showers, respectively, from the 
direction of 1ES~1959+650. In the following we use $\dot n_{\rm atmos} = 
0.5$~events/yr for AMANDA-II, and estimate $\dot n_{\rm atmos} = 4$~events/month 
for IceCube.
The number of observed events $N_{obs}$ with $N_{\rm obs} = 
N_{\rm AGN} + N_{\rm atmos}$
follows a Poissonian distribution
$$
P(N_{\rm obs},N_{\rm atm}) = \frac{N_{\rm atm}^{N_{\rm obs}}}{N_{\rm obs}!} 
\exp(-N_{\rm atm})\,\, ,
$$
and by appropriate summation of this distribution we can determine the event 
rate of AGN neutrinos required
for a signal of pre-specified significance.
Fig.~\ref{sig} gives the result for a $5 \, \sigma$ excess of 100~GeV -- 
100~TeV AGN neutrinos in a predetermined $2 \times 10^{-3}$~sr resolution 
element.
Obviously, AMANDA-II would need AGN flare exposures of order 
years for only a $5 \, \sigma$ signal, while the required
exposure time 
scale for IceCube is of order months. The duration of the 1ES~1959+650 
orphan flare
in June 2002 was only a fraction of an hour. Mkn~421 had
similar flare durations.
This indicates that many orphan flare states 
have to be co-added to produce meaningful signals, even
in the IceCube detector.

\section{\label{conclusion}Conclusions}

We presented calculations of the neutrino flux and $\nu_\mu$ event
rate from the expected neutrino flare that was accompanied
by the 
$\gamma$-ray orphan flare of 1ES~1959+650 in June 2002, within the 
recently proposed hadronic Synchrotron Mirror model.
In this model, 
neutrino production is the result of decaying charged mesons that 
are produced in hadronic interactions of photons and relativistic 
protons in the giant resonance region of the cross section,
with 
the target photon field built up through reflecting the primary 
flare synchrotron
emission at clouds at a distance of a few pc from 
the central engine. The duration of this $\nu$-outburst is of order 
$\sim 1/3$~hr, comparable to the orphan TeV-flare,
and should occur 
$\sim 15$~days after the primary TeV- and contemporaneously with the
orphan TeV-flare.

The expected neutrino flux level ranges between $\sim 8 \times 10^{-10}$
and $1.3 \times 10^{-8}$~cm$^{-2}$~s$^{-1}$
($>100$ GeV), depending on 
the cosmic background model and proton injection spectral indices
($\alpha_p = 2$ -- 2.5). Given the short duration of this $\nu$-flare, 
the expected rate in AMANDA-II
is of order $10^{-4} \, \nu_\mu$ events per 
hour and thus insufficient to explain the
recent $\nu$-event observed by 
AMANDA-II from the direction of 1ES~1959+650
at the time of the June 2002 
orphan TeV-flare. For a ground based neutrino observatory this signal at 
energies $\leq 10$~TeV has to compete
with the strong atmospheric neutrino 
background. An estimate of the required exposure time
for a $5 \, \sigma$
signal in AMANDA-like and IceCube-like instruments reveals
that AMANDA's 
capabilities do not allow to detect single $\nu$-flares within the 
HSM model, and
an accumulation of order 
~years of orphan flare states 
would be necessary to produce a meaningful signal.
This situation 
significantly improves with km$^3$ arrays like IceCube where
a $5 \, 
\sigma$ signal is expected by co-adding less than $\sim 2$ months of
flaring data.

\acknowledgments
AR acknowledges financial support by a Lise-Meitner fellowship.
MB acknowledges partial support by NASA through XMM-Newton GO 
grant no. NNG~04GF70G.

\newpage

\newpage

\newpage

\newpage

\end{document}